\documentclass[sigconf,nonacm]{acmart}
\usepackage[utf8]{inputenc}
\usepackage{xspace}
\usepackage{soul}
\usepackage{tabularx}
\usepackage{caption}
\usepackage{subcaption}
\usepackage{adjustbox}
\usepackage{longtable}
\usepackage{array}
\usepackage{wrapfig,lipsum,booktabs}
\usepackage{textcomp}
\usepackage{multicol}
\usepackage{multirow}
\usepackage{comment}
\usepackage{amsmath}
\usepackage{enumitem,ragged2e}
\usepackage{xcolor,colortbl}
\usepackage[most]{tcolorbox}
\usepackage{listings}
\usepackage{amsmath}      
\usepackage{listings}     
\usepackage{xcolor}       
\usepackage{graphicx}
\usepackage{algorithm}
\usepackage{algpseudocode}
\usepackage{calc}
\usepackage{hyperref}
\usepackage[utf8]{inputenc}
\definecolor{keywordcolor}{rgb}{0.6, 0, 0}  
\definecolor{commentcolor}{rgb}{0, 0.6, 0}  
\definecolor{stringcolor}{rgb}{0.2, 0.4, 0.8}
\definecolor{backgroundcolor}{rgb}{0.9,0.9,1.0}  

\AtBeginDocument{%
  }

\begin{document}


\title{Towards a HIPAA Compliant Agentic AI System in Healthcare}

\author{Subash Neupane}
\orcid{https://orcid.org/0000-0001-9260-3914}
\affiliation{%
  \institution{Mississippi State University}
  \city{Starkville}
  \state{Mississippi}
  \country{USA}
}
\email{sn922@msstate.edu}

\author{Sudip Mittal}
\affiliation{%
  \institution{Mississippi State University}
  \city{Starkville}
  \state{Mississippi}
  \country{USA}}
\email{mittal@cse.msstate.edu}

\author{Shahram Rahimi}
\affiliation{%
  \institution{University of Alabama}
  \city{Tuscaloosa}
  \state{Alabama}
  \country{USA}
}
\email{srahimi1@ua.edu}

\renewcommand{\shortauthors}{Neupane et al.}

\begin{abstract}


Agentic AI systems powered by Large Language Models (LLMs) as their foundational reasoning engine, are transforming clinical workflows such as medical report generation and clinical summarization by autonomously analyzing sensitive healthcare data and executing decisions with minimal human oversight. However, their adoption demands strict compliance with regulatory frameworks such as Health Insurance Portability and Accountability Act (HIPAA), particularly when handling Protected Health Information (PHI). This work-in-progress paper introduces a HIPAA-compliant Agentic AI framework that enforces regulatory compliance through dynamic, context-aware policy enforcement. Our framework integrates three core mechanisms: (1) Attribute-Based Access Control (ABAC) for granular PHI governance, (2) a hybrid PHI sanitization pipeline combining regex patterns and BERT-based model to minimize leakage, and (3) immutable audit trails for compliance verification. 

\end{abstract}

%
%



\maketitle

\section{Introduction}


Agentic AI system is an emerging paradigm in Artificial Intelligence (AI) where autonomous systems pursue complex goals with minimal human intervention \cite{shavit2023practices}. The integration of these systems powered by Large Language Models (LLMs) into healthcare workflows represents a paradigm shift in clinical decision making and administrative efficiency \cite{sudarshan2024agentic}. Unlike passive LLMs, Agentic AI systems can dynamically interact with Electronic Health Records (EHRs), synthesize multimodal patient data, and execute context-aware tasks such as generating clinical documentation, summarizing medical literature, or offering real-time diagnostic recommendations \cite{sudarshan2024agentic,10.1145/3709365,10825266}. 


\begin{figure}[htbp]
    \centering
    \includegraphics[scale=1.48]{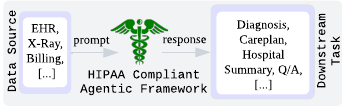}
    \caption{A graphical overview of the HIPAA-compliant Agentic framework.  [...] (used for brevity) indicates
that there are more data modalities and downstream tasks.}
        \label{fig: overview_figure}
\end{figure} 

However, the autonomous nature of these systems introduces critical risks to the security of Protected Health Information (PHI) as mandated by the Health Insurance Portability and Accountability Act (HIPAA) \cite{act1996health}. PHI is defined by the U.S. Department of Health and Human Services (HHS) in the HIPAA Privacy Rule (45 CFR \S\ 160.103) as ``\textit{Individually identifiable health information}'' that is transmitted or maintained in any form or medium (electronic, paper, or oral) by a covered entity or its business associates, and relates to \textit{past, present, or future physical or mental health or condition of an individual, provision of health care to an individual,} and \textit{past, present, or future payment for the provision of health care to the individual}. An example containing multiple PHI elements (name, birth date, medical record number, diagnosis, treatment information, visit date, and insurance identifier) that would require protection under HIPAA is shown in Fig. \ref{fig:phi}.
\begin{figure}[ht]
\vspace{-3mm}
    \centering 
    \begin{minipage}{8.4cm}
    
        \begin{tcolorbox}[enhanced,attach boxed title to top center={yshift=-1mm,yshifttext=-1mm},
            colback=blue!10!white,colframe=gray!90!black,colbacktitle=gray!80!black, left=0.1mm, right=0.5mm, boxrule=0.50pt]
            \footnotesize
            {\fontfamily{cmr}\selectfont
           \texttt{Patient Name: Barry Berkman
DOB: 01/15/1965
Medical Record Number: MRN-12345678
Diagnosis: Type 2 Diabetes
Medication: Metformin 500mg twice daily
Last Visit: 03/02/2025
Insurance ID: ABC123456789}
            }
        \end{tcolorbox}
    \end{minipage}    
    \vspace{-3mm}
    \caption{An example of PHI identifiers that requires protection  based on HIPAA \textit{Safe Harbor} rule \S\ 164.514(b)(2).}
   
    \label{fig:phi}
\end{figure}

HIPAA regulations require healthcare providers to implement technical, physical, and administrative safeguards to restrict access to PHI and ensure that only authorized users can interact with patient data. A critical component of these safeguards is access control, which governs who can view, modify, or transmit sensitive information. However, in the context of Electronic Health Records (EHR), where clinical notes (e.g. diagnoses, treatment plans, medical histories) are often stored as unstructured free text 
traditional access control mechanisms struggle to enforce granular protections. Unstructured data inherently contains sensitive identifiers (e.g. names, addresses) and complex clinical narratives, creating vulnerabilities that static access policies cannot fully mitigate. These risks are amplified with the integration of Agentic AI systems powered by LLMs in healthcare workflow. Without dynamic safeguards, LLM driven workflows risk inadvertently exposing PHI, memorizing sensitive details during training, or bypassing rigid access rules, violating HIPAA’s \textit{Minimum Necessary Standard} (\S\ 164.502(b)).

The current literature predominantly explores Agentic AI systems for specific downstream tasks, such as identifying cognitive concerns in clinical notes \cite{tian2025agentic}, generating medical reports \cite{sudarshan2024agentic}, and delivering clinical services \cite{reddy2025enabling}. Nevertheless, these systems demonstrate significant limitations in achieving comprehensive HIPAA regulatory compliance, thus constraining their potential for widespread clinical deployment. To address regulatory and compliance gaps in previous works, this work-in-progress paper introduces a novel HIPAA compliant Agentic AI framework that provides technical safegurad to protect PHI. Fig. \ref{fig: overview_figure} provides a high-level overview of our framework for several downstream tasks such as diagnosis prediction, care plans, clinical summary generation, radiology report generation, and question and answer. Our framework implements HIPAA mandates through three core mechanisms: Attribute-Based Access Control (ABAC) (\S\ 164.312(a)(1)), which dynamically restricts data access based on user roles, resource sensitivity, and environmental context; layered PHI sanitization (\S\ 164.514(b)(2)), applying dual redaction stages (pre- and post-inference) to minimize PHI exposure; and immutable audit trails (\S\ 164.312(b)), which log all access events and policy decisions for compliance verification.

The major contributions of this paper are:
\begin{itemize}
    \item An Agentic AI framework integrating ABAC, sanitization, and audit agents to enforce HIPAA compliance in healthcare workflows.
    \item A hybrid PHI sanitization pipeline that combines regex patterns and BERT-based model to mitigate PHI leakage meeting HIPAA \textit{Safe Harbor} and \textit{Expert Determination} de-identification rule. 
    \item Preliminary evaluation demonstrating the accuracy of PHI detection and the efficiency of the system.
\end{itemize}


\section{Background}
In this section, we provide background on HIPAA, explore Agentic AI systems in healthcare, and discuss the Access Controls.
\label{background}
\subsection{Health Insurance Portability and Accountability Act (HIPAA)}
\label{hippa_constraints}
HIPAA, enacted in 1996, is a landmark federal law in the United States designed to protect PHI, ensure continuity of health insurance coverage, and modernize healthcare data exchange. The regulatory framework of HIPAA, codified in Title 45 of the Code of Federal Regulations (CFR), imposes stringent requirements on covered entities (e.g. healthcare providers, insurers) and their business associates to protect patient privacy and secure health data. Its provisions are particularly critical in an era of EHR and advanced technologies, such as Artificial Intelligence (AI) and LLM, which process vast amounts of sensitive health data.

\subsubsection{Core Components of HIPAA}
HIPAA mandates comprise multiple regulatory rules,
each governing distinct facets of PHI management and oversight. 

    \textbf{Privacy Rule} (45 CFR \S\ 160 and 164, Subpart E) : The privacy rule establishes standards for the use and disclosure of PHI, defined as individually identifiable health information transmitted or maintained in any form (45 CFR \S\ 160.103). The key provisions include (i) \textit{Minimum Necessary Standard} which means covered entities must limit the access, use, or disclosure of PHI to the minimum necessary to achieve the intended purpose (45 CFR \S\ 164.502 (b)). (ii) \textit{Patient Rights}- individuals retain the right to access (\S\ 164.524), request amendments (\S\ 164.526) and obtain an accounting of disclosures (\S\ 164.528) of their PHI. (iii) \textit{Authorization Requirements} which entails written patient consent is mandatory for non-routine disclosures unrelated to treatment, payment or healthcare operations (45 CFR \S\ 164.508).

    \textbf{Security Rule} (45 CFR \S\ 164.302–318, Subpart C):
    The security rule establishes mandatory protections for electronic Protected Health Information (ePHI), requiring covered entities to implement a comprehensive framework across three core safeguard categories. \textit{Administrative Safeguards} focus on organizational policies, including Risk Analysis Requirements (\S\ 164.308(a)), workforce security training (\S\ 164.308(a)(5)), and access control protocols (\S\ 164.308(a)(4)). \textit{Physical Safeguards} address facility-level protections, mandating controlled facility access (\S\ 164.310(a)) and device/media disposal security (\S\ 164.310(d)). \textit{Technical Safeguards} enforce digital security through encryption standards (\S\ audit controls (\S\ 164.312(b)), and user identity verification mechanisms (\S\ 164.312(d)). Together, these layered requirements ensure ePHI remains confidential, integral, and available only to authorized personnel.

    \textbf{Breach Notification Rule} (45 CFR \S\ 164.400–414, Subpart D): 
    This rule requires covered entities to notify affected individuals, HHS, and, in some cases, the media following a breach of unsecured PHI. A breach is defined as unauthorized access, acquisition, or disclosure that compromises privacy (\S\ 164.402). Notifications must occur within 60 days of discovery (\S\ 164.404).

\subsection{Agentic AI Systems in Healthcare}
\label{usage_con_llm}

Agentic AI systems have transformative potential for healthcare delivery by enabling iterative, role-specialized workflows that improve diagnostic precision and operational efficiency \cite{tian2025agentic}. Agentic AI systems deploy modular agents to execute discrete tasks such as clinical summary generation or care plan interpretation with minimal human intervention. For example, an administrative agent could autonomously generate EHR updates, reducing documentation burdens linked to physician burnout 
, while a patient facing agent synthesizes care plans into layperson friendly summaries, mitigating risks of misinterpretation \cite{10825266}. Frameworks like MedInsight \cite{10.1145/3709365} exemplify agentic principles, where context-retrieval agents augment medical Q\&A systems with patient specific medical histories and authoritative knowledge, empowering patients and caregivers through personalized, targeted, and contextual medical recommendations. 
However, while current Agentic AI systems in healthcare excel in few downstream tasks such as cognitive concern detection \cite{tian2025agentic}, medical report generation \cite{sudarshan2024agentic}, or automated clinical services \cite{reddy2025enabling} they lack robust mechanisms to ensure end-to-end HIPAA compliance.

\subsection{Attribute-Based Access Control (ABAC)}

ABAC is a dynamic authorization framework that grants or restricts access to resources based on attributes associated with subjects (users or agents), resources (data or services), actions (operations performed), and environmental conditions (for example, time, location) \cite{hu2015attribute}. Unlike Role-Based Access Control (RBAC), which relies on static role assignments, ABAC evaluates contextual attributes in real time, enabling granular, policy-driven decisions. This flexibility makes ABAC particularly suited for complex systems such as Agentic AI workflows in healthcare, where access requirements vary dynamically across tasks and stakeholders.

\section{HIPAA Compliant Agentic AI Framework}
\label{approach}
\begin{figure*}[htbp]
    \centering
    \includegraphics[scale=1.1]{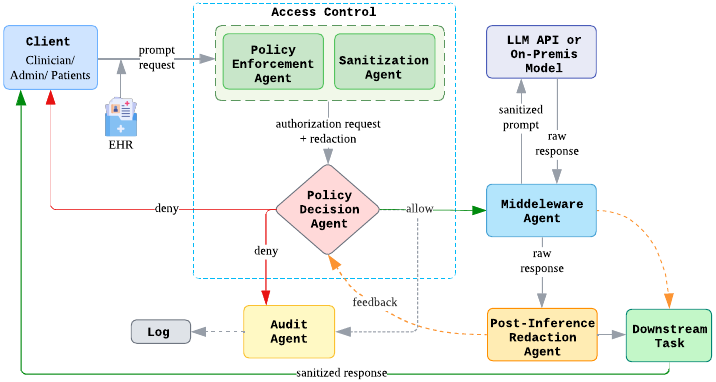}
    \caption{Architecture of the HIPAA-compliant Agentic AI Framework for Clinical Workflows. The system integrates dynamic Attribute-Based Access Control (ABAC), hybrid PHI sanitization, and immutable audit trails to enforce compliance across autonomous data interactions in healthcare settings.}
        \label{fig: system_architecture}
        \vspace{-4mm}
\end{figure*} 

The overarching goal of our framework is to ensure continuous, policy-driven HIPAA compliance for Agentic AI systems operating within clinical workflows such as medical report generation. This section details the framework’s components, which enforce compliance through context-aware policy specification, dynamic access control, hybrid PHI sanitization, and immutable audit trails, all designed to govern the autonomous decision-making inherent to Agentic AI. Figure \ref{fig: system_architecture} illustrates the high-level architecture, emphasizing how these mechanisms interact to secure sensitive healthcare data across all stages of Agentic AI systems.


\subsection{Access Control}
\label{auth_engine}

The core of our methodology leverages ABAC to dynamically govern access to PHI in Agentic AI systems. Unlike traditional role-based models, ABAC evaluates subject attributes (e.g., user roles), resource attributes (e.g., data sensitivity), action types (e.g., read/write), and environmental attributes (e.g., time, network security) to enforce least-privilege access. Policies are defined using first-order logic and enforced through a distributed architecture of policy agents.

Let:
\begin{itemize}
\small
    \item \( S = \{s_1, \dots, s_n\} \): Subjects with \( A_s = \{\text{role, department, clearance}\} \)
    \item \( O = \{o_1, \dots, o_m\} \): Objects with \( A_o = \{\text{type, sensitivity, owner}\} \)
    \item \( A = \{a_1, \dots, a_p\} \): Actions
    \item \( E = \{e_1, \dots, e_k\} \): Environmental attributes
\end{itemize}

Authorization is granted iff:
\begin{equation}
\text{Authorize}(s, o, a) \iff \bigwedge_{i} \phi_i(A_s, A_o, a, E),
\end{equation}
where $\phi_i$ are predicates combining attributes. For example, let $\phi_1$ be defined as: {role$(s)$ = cardiologist $\land$ sensitivity$(o) \leq 2$ $\land$ $(a$ = read) $\land$ time$(e) \in [8, 18]$}. Obligations (e.g., logging, sanitization) are enforced as part of policy decisions such as:
\begin{equation}
\left(\bigwedge_{i} \phi_i(A_s, A_o, a, E)\right) \implies \text{Authorize}(s, o, a) \land \text{Enforce}(\mathcal{O})
\end{equation}
where $\mathcal{O} = \{\text{Log}(s, o, t), \text{Sanitize}(o)\}$.  

Policies are codified in an XACML-like syntax that maps directly to the formal model. For example, the rule for a cardiologist described above translates to Listing \ref{lst:abac}.
\begin{lstlisting}[language=XML, caption={ABAC Policy for Cardiac Data Access}, label=lst:abac]
<Policy>
  <Target>
    <SubjectAttributes>
      <Attribute Name="role" Value="cardiologist"/>
    </SubjectAttributes>
    <ResourceAttributes>
      <Attribute Name="data_type" Value="cardiac"/>
    </ResourceAttributes>
    <Action>Read</Action>
  </Target>
  <Condition>
    <EnvironmentAttribute Name="time" Value="8<=t<=18"/>
  </Condition>
  <Obligations>
    <Obligation>log_access</Obligation>
    <Obligation>sanitize_phi</Obligation>
  </Obligations>
</Policy>
\end{lstlisting}

The sanitization agent enforces PHI de-identification and redaction as part of obligations, ensuring HIPAA’s \textit{Minimum Necessary} and \textit{De-Identification rules}. We utilize a hybrid approach combining rule-based regex patterns and a BERT-based model pipeline to detect and redact PHI. Regular expressions target structured identifiers (e.g., Social Security Numbers: \texttt{$\backslash$d\{3\}--$\backslash$d\{2\}--$\backslash$d\{4\}, Medical Record Numbers: [A-Z]$\backslash$d\{6\}}) with deterministic pattern matching, ensuring efficient removal of standardized PHI formats. For unstructured text, a BERT model fine-tuned in clinical corpora (e.g., MIMIC-IV discharge notes) \cite{johnson2023mimic} identifies contextual PHI, such as patient names, diagnoses or provider details, which lack rigid syntactic patterns. This approach complies with HIPAA's \textit{expert determination} 45 CFR \S\ 164.514(b)(1) for PHI redaction. For example, the input ``John Smith, 55, diagnosed with NSCLC in 2022'' is sanitized to ''[PatientName], [Age], diagnosed with [Condition] in [Year]'' by masking regex-matched ages/years and BERT-recognized entities.

The Policy Decision Agent (PDA) is responsible for evaluating access requests against predefined security and privacy policies. Acting as a decision-making engine, the PDP analyzes contextual attributes such as user roles, resource types, actions, and environmental conditions to determine whether access should be granted or denied. A high-level overview of the PDP algorithm is shown in Algorithm \ref{algo:pdp}.
\vspace{-3mm}
\begin{algorithm}

\caption{Policy Decision Agent}
\label{algo:pdp}
\begin{algorithmic}[1]
\Function{EvaluateRequest}{$s, o, a, E$}
    \State Load policies for $(A_s, A_o, A_a, E)$
    \For{each policy $\in$ policies}
        \If{MatchAttributes($A_s$, $A_o$, $A_a$, $E$)}
            \State \Return (ALLOW, $\mathcal{O}$) \Comment{With obligations}
        \EndIf
    \EndFor
    \State \Return DENY
\EndFunction
\end{algorithmic}
\end{algorithm}

\vspace{-4mm}

\subsection{Middleware Agent}

LLMs in Agentic systems operating as on-premises models or third-party APIs require continuous attribute-based governance. For API calls  a Business Associate Agreement (BAA) is mandatory under HIPAA. In such case, BAA compliance must be included as environment attribute such that: 
\begin{equation}
\text{Authorize}(s, o, a) \iff \phi_i(A_s, A_o, A_a, E) \land \text{BAA}_{valid}({\text{API}}_p)
\end{equation}

where, $BAA_{valid}$ is a boolean check confirming an active BAA with the third-party API provider and $API_{p}$ is an attribute identifying the external service (e.g., "AWS", "Azure OpenAI").

The middleware agent intercepts all requests/responses to/from the LLM, enforcing dynamic policy evaluation through three core mechanisms such as \textit{session attribute tracking, stateful policy reevaluation,} and \textit{conversation context analysis}.

    \textbf{Session Attribute Tracking}: It maintains a real-time session state through three critical attributes:
    \begin{itemize}
        \item $\mathit{user\_role}$ (e.g., ``nephrologist'', ``billing\_specialist'')
        \item $\mathit{consent\_status}$ (active/revoked)
        \item $\mathit{phi\_access\_count}$ (number of PHI elements accessed)
    \end{itemize}
    
    \textbf{Stateful Policy Reevaluation}: Triggers PDA reauthorization when session attributes change. An example of consent revocation during session is presented in Listing \ref{lst:abacd}. 
   \begin{lstlisting}[language=xml, caption={ Consent Revocation Policy}, label=lst:abacd]  
<Policy PolicyId="MW-Revoke">  
  <Condition>  
    <AttributeMatch AttributeId="consent_status" Value="revoked"/>  
  </Condition>  
  <Obligations>  
    <Obligation>TERMINATE_SESSION</Obligation>  
    <Obligation>DELETE_CACHED_PHI</Obligation>  
  </Obligations>  
</Policy>  
\end{lstlisting}  
    
    \textbf{Conversation Context Analysis}:
    We prevent incremental disclosure of PHI by tracking cumulative risk across a user session. For this, we calculate risk scores based on factors such as \textit{PHI Sensitivity} and \textit{Access frequency}. Each type of PHI (e.g., SSN, diagnosis) is assigned a sensitivity level (1 = low, 5 = high), and how often PHI is accessed in the current session. The risk score is calculated as:
\[
\scalebox{0.9}{$\text{risk\_score} = \left(\sum \text{sensitivity of PHI }\right) \times \left(\text{number of PHI accesses}\right)$}
\]
A cardiologist might be allowed a higher risk threshold (e.g., 20) than a billing clerk (e.g., 10) before access is blocked.


\subsection{ Post-Inference Redaction Agent}
This agent re-sanitizes Agentic AI outputs to address residual PHI leakage. Similarly to the sanitization agent discussed earlier (see Section \ref{auth_engine}) it uses both HIPAA's \textit{Safe Harbor} and \textit{Expert determination} de-identification techniques. We use PDA generated obligations for attribute-driven redaction. Table \ref{tab:redaction_obligation} presents an overview of the redaction obligation.
    \begin{table}[h]
    \vspace{-4mm}
        \centering
        \caption{An overview of attribute driven redaction.}
        \vspace{-4mm}
        \begin{tabular}{ll}
        \hline
        \textbf{Obligation} & \textbf{Redaction Action} \\ \hline
        REDACT\_ALL & Replace all PHI with [REDACTED] \\ 
        REDACT\_DEMO & Remove demographics (names, DOB) \\ 
        MASK\_CODES & Replace ICD-10 codes with category tags \\ \hline
        \end{tabular}
        
        \label{tab:redaction_obligation}
        \vspace{-3mm}
    \end{table}

 Following post-inference redaction, the redacted response is disseminated through two paths to balance usability and compliance. First, the sanitized output is delivered to the end-user, ensuring only HIPAA-compliant data is released per the \textit{Minimum Necessary Standard} (\S\ 164.502(b)). Concurrently, both raw and sanitized responses are archived in the Audit Agent’s cryptographically secured ledger (see \ref{audit_agent}), creating an immutable, tamper-evident audit trail. This dual-path approach enforces accountability while fulfilling HIPAA’s 6-year retention mandate (45 CFR \S\ 164.316), ensuring retrospective compliance verification without obstructing real-time clinical or administrative workflows.


\subsection{Audit Agent}
\label{audit_agent}

The audit agent implements dual logging architecture based on the National Institute of Standards and Technology (NIST) 800-66r2 \cite{guide2024implementing} such as \textit{interaction logs} and \textit{decision logs}. The former records sanitized user queries, policy decisions, and redaction actions. Both raw LLM outputs and sanitized versions are stored for forensic investigations. The latter is an immutable ledger of access decisions (allow/deny) secured via cryptographic hashing to prevent tampering.

 \section{Preliminary Results}
 \label{results}

 \subsection{Dataset}

 We utilize Medical Information Mart for Intensive Care (MIMIC-IV) dataset \cite{johnson2023mimic}, a publicly available Electronic Health Record (EHR) repository from Beth Israel Deaconess Medical Center, accessible via PhysioNet. This dataset comprises more than 109,000 Emergency Department (ED) visits, each record including emergency stay diagnosis codes ICD-9 or ICD-10, chief complaints, at least one radiology report, and discharge summary. A summary of data statistics is presented in Table. \ref{tab:dataset_stats_discharge_note}.
\begin{table}[h]
\vspace{-4mm}
    \centering
    \caption{Dataset Statistics}
    \vspace{-3mm}
    \resizebox{\columnwidth}{!}{%
    \begin{tabular}{lcccc}
        \toprule
        \textbf{Item} & \textbf{Samples} & \textbf{Training} & \textbf{Validation} & \textbf{Testing}  \\
        \midrule
        Admissions & 109,168 & 68,785 & 14,719 & 10,962\\
        Discharge Summaries & 109,168 & 68,785 & 14,719 & 10,962  \\
        Radiology Reports & 409,359 & 259,304 & 54,650 & 40,608\\
        ED Stays \& Chief Complaints & 109,403 & 68,936 & 14,751 & 10,985  \\
        ED Diagnoses & 218,376 & 138,112 & 29,086 & 21,764 \\
        \bottomrule
    \end{tabular}%
    }
    \label{tab:dataset_stats_discharge_note}
    \vspace{-2mm}
\end{table}
We utilize discharge summary as our dataset to evaluate the effectiveness of the PHI sanitization technique described in Section \ref{auth_engine}. These summaries come with prior de-identification and anynomization as shown in Fig. \ref{fig:data_sample} where the PHI entities have been replaced and redacted with dashes. We first augment the redacted PHI with synthetic PHI using 0-shot inference using LLAMA 3.2. The augmented data are then validated with our sanitization and post-inference redaction agent. 

\begin{figure}[ht]
\vspace{-2mm}
    \centering 
    \begin{minipage}{8.4cm}
    
        \begin{tcolorbox}[enhanced,attach boxed title to top center={yshift=-1mm,yshifttext=-1mm},
            colback=blue!10!white,colframe=gray!90!black,colbacktitle=gray!80!black, left=0.1mm, right=0.5mm, boxrule=0.50pt]
            \footnotesize
            {\fontfamily{cmr}\selectfont
           \texttt{\textbf{Name}: ---- \textbf{Unit No:} ---- \textbf{Sex:} F \textbf{DOB:} ---- \textbf{Admission Date:} ---\\ Discharge Date ----\\
           \textbf{Service}: Medicine,  \textbf{Allergies}: Percocet\\
        \textbf{Chief Complaint}: Abdominal fullness and discomfort.\\
        \textbf{History of Present Illness:} ---- with HIV (on HAART), COPD, and cirrhosis complicated by ascites and hepatic encephalopathy [...]
        \textbf{Imaging:}---- CXR- No acute cardiopulmonary abnormality.----RUQ US  1. Extremely coarse and nodular [...]}
            }
        \end{tcolorbox}
    \end{minipage}    
    \vspace{-3mm}
    \caption{An example of unstructured discharge summary. --- represent anynomized PHI entities and [...] is used for brevity of textual contents.}
    \label{fig:data_sample}
  \vspace{-3mm}
\end{figure}

 \subsection{Evaluation}
 We present our preliminary findings on three key dimensions including PHI sanitization accuracy, policy enforcement efficiency, and risk mitigation. Table \ref{tab:sanitization} presents an overview of PHI sanitization accuracy. Our hybrid approach demonstrated superior performance, particularly in handling unstructured clinical narratives. The BERT model achieved 96.5\% recall for contextual PHI (e.g., diagnoses), while regex maintained 100\% precision for structured identifiers.

 \begin{table}[htbp]
\centering
\vspace{-2mm}
\caption{PHI Redaction Performance Comparison (N=500 notes, 2,350 PHI instances)}
\vspace{-4mm}
\label{tab:sanitization}
\begin{tabular}{@{}lccc@{}}
\toprule
\textbf{Metric} & \textbf{Regex-Only} & \textbf{BERT-Only} & \textbf{Hybrid} \\ 
\midrule
Precision & 98.2\% & 92.1\% & \textbf{99.4\%} \\
Recall & 67.3\% & 89.8\% & \textbf{97.6\%} \\
F1-Score & 79.8\% & 90.9\% & \textbf{98.4\%} \\
Residual PHI & 32 & 24 & \textbf{3} \\
\bottomrule

\end{tabular}
\vspace{-4mm}
\end{table}

The policy enforcement efficiency was evaluated by evaluating the decisions made by PDA. For this, we we evaluated 200 simulated access requests across multiple roles (e.g., cardiologist, billing clerk) and data sensitivity levels. Key performance metrics included \textit{ABAC Policy Matching Accuracy}, \textit{Decision Latency}, and \textit{Risk Threshold Enforcement}. The PDA demonstrated 99.1\% accuracy in dynamically granting/denying access based on contextual attributes including user roles and temporal constraints. The system maintained an average decision time of 12.3ms (SD = 2.1~ms), satisfying real-time requirements for clinical workflow integration.  Sessions exceeding role-specific risk thresholds (cardiologist: 20, billing clerk: 10) were terminated with 100\% reliability, preventing potential \textit{PHI} over-exposure.  Furthermore, consent revocation events (as implemented in Listing~\ref{lst:abacd}) resulted in instantaneous session termination and deletion of cached PHI data, demonstrating effective compliance with HIPAA's Right to Revoke provision (\S\ 164.508(b)(5)).

\section{Discussion, Futurework \& Conclusion}

Our HIPAA-compliant Agentic AI framework addresses the critical challenge of securing autonomous workflows in healthcare by integrating three core mechanisms: (1) dynamic Attribute-Based Access Control (ABAC) to enforce granular, context-aware permissions over unstructured EHR data, (2) a hybrid PHI sanitization pipeline combining regex and BERT-based model to minimize leakage in free-text narratives, and (3) immutable audit trails to ensure accountability under HIPAA’s retention mandates. While our synthetic PHI evaluation followed HIPAA \textit{Safe Harbor} and \textit{Expert Determination} guidelines, real-world deployment may reveal edge cases in free-text documentation patterns. Preliminary results demonstrate the the effectiveness of our framework. By embedding HIPAA’s \textit{Minimum Necessary Standard} (\S\ 164.502(b)) into every stage of data interaction, our framework advances the responsible deployment of Agentic AI systems in clinical settings. Future work will extend these safeguards to multi-modal data (e.g., imaging, genomics) and adversarial scenarios, ensuring scalability as healthcare workflows increasingly adopt Agentic AI systems.

\section*{Acknowledgment}

This work was supported by the Predictive Analytics and Technology Integration (PATENT) Laboratory at the Department of Computer Science and Engineering, Mississippi State
University.

\bibliographystyle{unsrt}
\bibliography{sample-base}

\appendix

\end{document}